\documentclass[5p,longtitle]{elsarticle}

\usepackage{graphicx}
\usepackage{txfonts}
\usepackage[table]{xcolor}
\usepackage{natbib}
\bibpunct{(}{)}{;}{a}{}{,}
\usepackage[english]{babel}

\usepackage{hyperref}
\hypersetup{colorlinks=true, citecolor=blue, linkcolor= magenta}
\usepackage{multirow}
\usepackage{makecell}
\usepackage{natbib}
\bibpunct{(}{)}{;}{a}{}{,}
\def\araa{ARA\&A}%
\def\apj{The Astrophysical Journal}%
\def\apjl{The Astrophysical Journal, Letters}%
\def\apjs{ApJS}%
\def\aap{Astronomy \& Astrophysics}%
\def\mnras{MNRAS}%
\def\na{New A}%
\def\nar{New A Rev.}%
\def\pasj{PASJ}%
\def\physrep{Phys.~Rep.}%
\def\procspie{Proc.~SPIE}%

\usepackage[nolist,printonlyused]{acronym}

\begin{acronym}[TDMA]
 \acro{LMXB}{Low mass X-ray binary}
 \acrodefplural{LMXB}{Low mass X-ray binaries}
 \acro{LMXBT}{low mass X-ray binary transient}
 \acro{AMXP}{Accreting millisecond X-ray pulsar}
 \acro{RXTE/PCA}{the Rossi X-ray timing explorer/proportional counter array}
 \acro{RXTE}{the Rossi X-ray timing explorer}
 \acro{MAXI}{monitor of all sky X-ray image}
 \acro{HMXB}{high mass X-ray binary}
 \acrodefplural{HMXB}{High mass X-ray binaries}
 \acro{Swift/XRT}{the Swift gamma-ray burst mission/X-ray telescope}
 \acro{Swift}{the Swift gamma-ray burst mission}
 \acro{XSPEC}{an X-ray spectral fitting package}
 \acro{FRED}{fast-rise-exponential-decay}
 \acro{DIM}{the disk instability model}
 \acro{LIS}{low-intensity-state}
 \acro{SXT}{soft X-ray transient}
 \acro{HS}{high/soft state}
 \acro{LH}{low/hard state}
 \acro{PCU}{proportional counter unit}
 \acro{ISS}{international space station}
 \acro{MSP}{millisecond pulsar}
 \acro{NS}{neutron star}
 \acro{Insight-HXMT}{Insight -- hard X-ray modulation telescope}
 \acro{HXMTDAS}{Insight-HXMT data analysis software}
\end{acronym}

\begin{document}
\begin{frontmatter}

   \title{Diagnostic of the spectral properties of Aquila X-1 by Insight-HXMT snapshots during the early propeller phase}

%
\author[1,2]{C.~G\"ung\"or}
\author[1]{M. Y.~Ge}
\author[1]{S.~Zhang}
\author[1,3]{A.~Santangelo}
\author[1,4,5]{S. N.~Zhang}
\author[1]{F. J.~Lu}
\author[1,4]{Y.~Zhang}
\author[1]{Y. P.~Chen}
\author[1]{L.~Tao}
\author[1]{Y. J. Yang}


% ...

\address[1]{Key Laboratory of Particle Astrophysics, Institute of High Energy Physics, Chinese Academy of Sciences, 19B Yuquan Road, Shijingshan District, 100049, Beijing, China}
\address[2]{Sabanc{\i} University, Faculty of Engineering and Natural Science, Orhanl{\i} $-$ Tuzla, 34956, \.{I}stanbul, Turkey}
\address[3]{Institut f\"ur Astronomie und Astrophysik, Universit\"at T\"ubingen, Sand 1, D 72076 T\"ubingen, Germany}
\address[4]{University of the Chinese Academy of Sciences, 100049, Beijing , China}
\address[5]{National Astronomical Observatories, Chinese Academy of Sciences, Beijing 100012, China}
\address[6]{Department of Astronomy, Beijing Normal University, Beijing 100088, People’s Republic of China}
\address[7]{Department of Physics, Tsinghua University, Beijing 100084, China}


\date{Accepted version 1.0, Month day, 2019}


\begin{abstract}

We study the 2018 outburst of Aql X-1 via the \ac{MAXI} data. 
We show that the outburst starting in February 2018 is a member of \textit{short-low}
class in the frame of outburst duration and the peak count rate although the outburst morphology is 
slightly different from the other 
\ac{FRED} type outbursts with a milder rising stage.
We study the partial accretion in the 
weak propeller stage of Aql X-1 via the \ac{MAXI} data of the 2018 outburst.
We report on the spectral analysis of 3 observations  of Aquila X-1 obtained by Insight -- hard X-ray modulation telescope
(Insight-HXMT) during the late decay stage of the 2018 outburst.
We discuss that the data
taken by Insight-HXMT is just after the transition to the weak propeller stage.
Our analysis shows the necessity of a comptonization component to take into account the existence of 
an electron cloud resulting photons partly up-scattered.

\end{abstract}


\end{frontmatter}

\section{Introduction}
\acp{LMXB} are systems containing a black hole (BH) or a \ac{NS} and a low mass 
companion ($M_{\mathrm c} \lesssim 1\,M_{\odot}$). The mass accretion mechanism in these systems is Roche 
lobe overflow \citep{fra+02}. The low mass companion fills its Roche lobe during its evolution and the material is being 
transferred from the first Lagrange point to the Roche lobe of the compact object. Since the 
transferring material has angular momentum, it creates an 
accretion disk around the compact object instead of falling onto its surface directly \citep{pri72}.
\acp{LMXB} may possible be incubators of millisecond pulsars.
Accretion onto NS can be the reason of conversion from the slow 
rotating NS with high magnetic field to the fast spinning and low magnetic field NS, so called recycling 
scenario \citep{bhat+91, tau+06}.

In the case of that the compact object is a NS, the disk material interacts with the magnetic field --the 
disk magnetosphere interaction--. The disk magnetosphere interaction shows different stages according to 
the location of the inner radius of disk:
\begin{itemize}
\item[\textit{(i)}] \textit{The accretion stage}: If the inner radius of the disk is closer to the \ac{NS} 
than the corotation radius\footnote{The radius that the angular velocity of the star is equal to the Keplerian 
angular velocity.}, all of the material reaching to the inner layers is being transferred onto the NS.
\item[\textit{(ii)}] \textit{The propeller stage}: The inner radius of the disk is located withinside the light 
cylinder\footnote{The cylinder centred on the pulsar and aligned with the rotation axis at whose radius, the 
corotating speed equals the speed of light.} and close to the  corotation radius \citep{ill75},
the material reaching to inner layers is dispersed for an ideal thin disk (full propeller).
If the inner layers of the disk have a scale height then a fraction of the material
can reach onto the poles \citep[weak propeller,][]{eks11} until the inner layers shrink back and the system becomes full propeller.
\item[\textit{(iii)}] \textit{The radio pulsar stage}: If the inner radius of the disk is located out of 
the light cylinder, the system acts as an isolated neutron star with no accretion.
\end{itemize}
A \ac{LMXB} completes these disk magnetosphere interaction stages along its billion years evolution.
\acp{AMXP} constitute a subset of \acp{LMXB} in which X-ray pulsations are 
detected, and they are peerless sources to study the evolution of \acp{LMXB} since they 
show disk magnetosphere interaction stages in an 
observable duration.

Aql X-1 is an NS-\ac{LMXB} accompanying a K4 $\pm$ 2 main sequence star,
rotating with the period of 19~h in a 36$^\circ$ $-$ 47$^\circ$ inclined orbit \citep{mata17}.
Aql X-1 is also classified as an \ac{AMXP} \citep{koy+81} 
with its intermittent pulsations with the pulse frequency of 550.27 Hertz detected only for 150 s overall 23 years of observations 
\citep{cas+08} in which the detected pulse frequency is consistent with the burst oscillations \citep{cas+08}.  
Aql X-1 is classified as \ac{SXT} showing cyclic outburst almost each year in its X-ray light curve. The X-ray Luminosity of the source in the quiescent state is
$L_{\rm   X}  \approx   10^{33}$~erg~s$^{-1}$  \citep{verb+94} while the outburst peak luminosity can exceed
$L_{\rm X}  \approx 10^{37}$~erg~s$^{-1}$ \citep{cam+13}.
\citet{jon04} reported the distance of the source as 4.4$-$5.9 kpc using the burst peak flux from \ac{RXTE} data.

Although Aql X-1 is one of the most studied source,
studies on this source may still shed light on a lot of open problems such as 
the physical origin of the intermittency of the pulsations.
The effect of comptonization to smear out pulsations has been argued in \citet{gogus07} 
who has not found any confirmation of such mechanism in three sources (GX 9+1, GX 9+9 and Sco X-1).
As they mentioned the sample set must be enlarged to generalize the outcome.
The observations, not only in outburst but also in
quiescent state, may allow us to study the time evolution of the crust temperature
and model crust cooling scenario. In Aql X-1 case, the outbursts are very frequent and 
there is no enough time between two outbursts to reach crust$-$core equilibrium \citep{ootes18}.
Spectral and temporal studies with Insight-HXMT data may play key role to address these open questions on \acp{LMXB} as well as the other X-ray space missions.

In this work, we present the spectral analysis outcome of the Insight-HXMT data.
We explain the details of the data reduction and present
our results of the spectral analysis in \autoref{obs}.
In \autoref{discuss}, we discuss the 2018 outburst in the frame of  outburst 
classification suggested by \citet{gungor+14} and
the partial accretion in the propeller stage \citep{gungor+17b}.
We also present the discussion on  the spectral outcome of Insight-HXMT data in \autoref{discuss}.
Finally, we conclude our study in \autoref{conc}.


\section{Observation and data analysis}
\label{obs}

\begin{figure} [!t]
\centering
  \includegraphics[angle=0,width=0.47\textwidth]{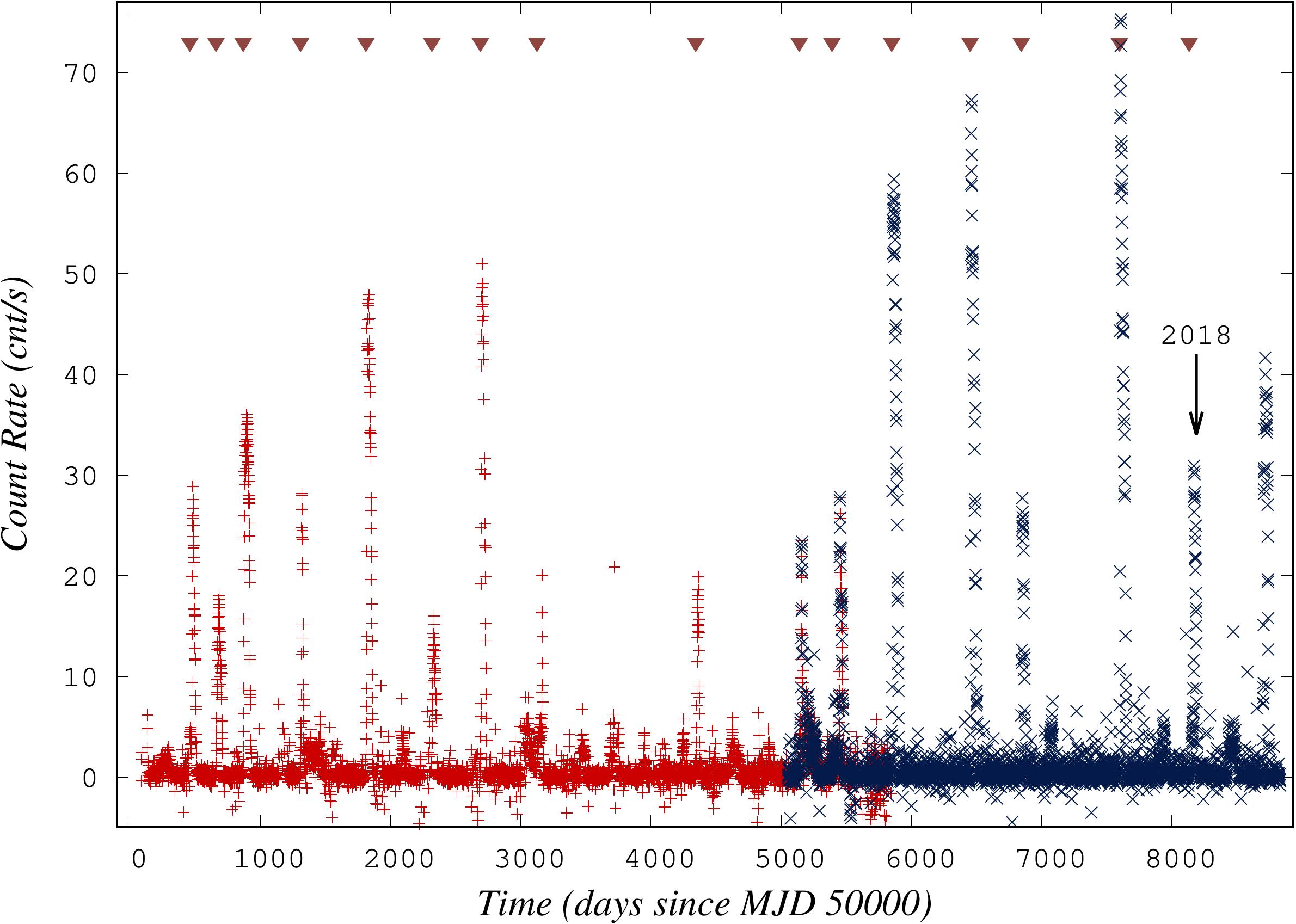}
     \caption{The long term light curve of Aql X-1 since 1996, 
     using the data from the ASM (dark red pluses) and the \ac{MAXI} (dark blue crosses).
     The count rates are given in the ASM level in which the \ac{MAXI} count rates are calibrated as explained in \autoref{obs}.
     The brown upside-down triangles show the \ac{FRED} type outbursts 
     and the black arrow indicates the 2018 outburst whose decay stage has been partly observed by Insight-HXMT.}
         \label{all}
\end{figure}

We show the 23 year light curve of Aql X-1 in \autoref{all} using the data taken 
by All-Sky Monitor \citep[ASM;][]{asm95}\footnote{\url{http://xte.mit.edu/ASM_lc.html}}
mounted on \ac{RXTE}
and the \ac{MAXI}\footnote{\url{http://maxi.riken.jp/}} \citep{mat+09} mounted on the \ac{ISS}. 
The \ac{MAXI} count rates are multiplied by 21.5 to calibrate to the ASM level 
by using the peak count rates of the 2009 and the 2010 outbursts which were jointly observed.
We used the one day binned \ac{MAXI} data in the energy range of $2.0-20.0$~keV 
to study the outburst starting in February 2018.
We obtained the hardness values during the 2018 outburst using the ratio of 
the \ac{MAXI} count rates in $2.0-10.0$~keV to the ones in $10.0-20.0$~keV.
The \ac{MAXI} light curve and the hardness evolution are used to classify the 2018 outburst
and to obtain the fastness parameters\footnote{A ratio of the angular velocity 
of the star to the Keplerian angular velocity at the inner disk radius $\omega_{\ast} \equiv \Omega_{\ast} / \Omega_{\mathrm K}(R_{\rm in})$.} for a given time using the method presented in \autoref{discuss}.

\begin{table*}[!t]
\begin{footnotesize}
\caption{Some properties of the Insight-HXMT observations and the 
best fit parameters of \textit{blackbody + power law} model (\textit{model~I}),
\textit{blackbody + disk blackbody} model (\textit{model~II}), 
\textit{blackbody + comptonization} model (\textit{model~III}), 
\textit{disk blackbody + comptonization} model (\textit{model~IV}), and 
\textit{blackbody  + disk blackbody + comptonization} model (\textit{model~V}).}
\begin{center}
\begin{tabular}{c|c|ccc}
\multicolumn{2}{c}{ObsID} & P011466801001 & P011466801002 & P011466801003 \\
\hline
\hline
\multicolumn{2}{c|}{Date (UTC)} & 2018/3/19 19:09:24 & 2018/3/20 19:01:00 & 2018/3/22 06:00:39 \\
\multicolumn{2}{c|}{Date (MJD)} & $58196.798937$ & $58197.793104$ & $58199.251194$ \\
\multicolumn{2}{c|}{Observation Mode} & Pointing & Pointing & Pointing \\
\multicolumn{2}{c|}{Exposure Time (ks)} & $10$ & $10$ & $10$ \\
\hline
\multirow{2}{*}{Total Count$^a$}	& LE	& $29371463$ & $38538779$ & $41110899$ \\
								& ME	& $116103644$ & $115698283$ & $159916052$ \\
								\hline
\multirow{2}{*}{Effective Count$^b$}	& LE	& $243040$ & $214303$ & $120770$ \\
								& ME	& $3654838$ & $3706650$ & $2357891$ \\
								\hline
\multicolumn{2}{c|}{$ \omega_{\ast}^{c}$} & $1.06$  & $1.07$  & $1.09$  \\
\hline 
\hline
\multirow{4}{*}{\makecell{Model I \\ bb+po}} 	
							& $kT_{bbody}$ (keV) 	& $4.64 \pm{0.50}$ 		& $8.36 \pm{1.76}$ 		& $11.91 \pm{4.50}$ \\
							& $\Gamma$				& $2.58 \pm{0.11}$ 		& $2.51 \pm{0.10}$ 		& $2.65 \pm{0.21}$ \\
							& Hardness$^{d}$ 		& $0.143 \pm{0.014}$ 	& $0.140 \pm{0.021} $ 	& $ 0.126 \pm{0.037} $ \\
							& $\chi^2/d.o.f.$ 		& $783/924$ 			& $824/924$ 			& $748/835$ \\
\hline
\multirow{4}{*}{\makecell{Model II \\ bb+diskbb}} 	
							& $kT_{bbody}$ (keV) 	& $3.62 \pm{0.24}$ 		& $5.13 \pm{0.59}$ 		& $6.33 \pm{1.45}$ \\
							& $kT_{diskbb}$ (keV) 	& $0.87 \pm{0.04}$ 		& $1.00 \pm{0.05}$ 		& $1.03 \pm{0.12}$ \\
							& Hardness$^{d}$		& $0.180 \pm{0.017}$	& $0.183 \pm{0.032} $ 	& $0.122 \pm{0.094} $ \\
							& $\chi^2/d.o.f.$ 		& $854/924$ 			& $831/924$ 			& $740/835$ \\
\hline
\multirow{4}{*}{\makecell{Model III \\ bb+compTT($T_{Wien}=kT_{bb}$)}} 	
							& $kT_{bbody}$ (keV) 	& $0.37 \pm{0.04}$ 		& $0.55 \pm{0.03}$ 		& $0.69 \pm{0.07}$ 		\\
							& $\tau$ 				& $2.63 \pm{0.33} $ 	& $4.99 \pm{0.87}$ 		& $6.94 \pm{2.08} $ \\
							& Hardness$^{d}$		& $0.255 \pm{0.013}$	& $0.185 \pm{0.021}$	& $0.103 \pm{0.013}$	\\
							& $\chi^2/d.o.f.$ 		& $781/924$ 			& $852/924$ 			& $732/835$ 			\\
\hline
\multirow{4}{*}{\makecell{Model IV \\ diskbb+compTT($T_{Wien}=kT_{diskbb}$)}}
							& $kT_{diskbb}$ (keV) 	& $0.46 \pm{0.06}$ 		& $0.78 \pm{0.07}$ 		& $0.98 \pm{0.11}$ 		\\
							& $\tau$ 				& $2.74 \pm{0.40} $ 	& $5.38 \pm{0.94}$ 		& $7.64 \pm{2.70} $ \\
							& Hardness$^{d}$		& $0.127 \pm{0.013}$	& $0.148 \pm{0.022}$	& $0.080 \pm{0.042}$	\\
							& $\chi^2/d.o.f.$ 		& $780/924$ 			& $835/924$ 			& $730/835$ 			\\
\hline 
\multirow{6}{*}{\makecell{Model V \\ bb+diskbb+compTT($T_{Wien}=kT_{bb}$)}}
					& $kT_{bbody}$		 			& $0.97 \pm{0.06}$ 		& $0.74 \pm{0.05}$ 		& $0.76 \pm{0.08}$ \\
					& $kT_{diskbb}$ 				& $0.48 \pm{0.01}$ 		& $0.29 \pm{0.01}$ 		& $0.38 \pm{0.04}$ \\
					& $\tau$ 						& $3.52 \pm{0.20} $ 	& $5.66 \pm{0.54}$ 		& $7.35 \pm{1.69} $ \\
					& $Flux^d~(10^{-9} erg/cm^2/s)$	& $0.413 \pm{0.009}$	& $0.263 \pm{0.005}$	& $0.077 \pm{0.005}$ \\
					& Hardness$^{e}$				& $0.119 \pm{0.011} $ 	& $0.138 \pm{0.009} $ 	& $0.074 \pm{0.019} $ \\
					& $\chi^2/d.o.f.$ 				& $770/922$ 			& $815/922$ 			& $727/833$ \\
							
\hline
\end{tabular}
\end{center}
\footnotesize{ 
\begin{flushleft}  $^{a}$ The count rates are the full counts obtained from the detector itself 
before background subtraction and good time interval corrections.\\
$^{b}$ The effective count rates are the total counts after good time interval corrections before background subtraction.\\
$^{c}$ The values of the fastness parameter are calculated using the $L_c$ obtained from 
the application of the method and using \autoref{fast2} explained in \autoref{prop} .\\
$^{d}$ Flux values are unabsorbed and calculated in the energy range of 2--20~keV using \textit{cflux} task in xspec.\\
$^{e}$ Hardness parameters are obtained using the unabsorbed flux ratio of two different energy ranges; $F(10.0-20.0~{\rm keV})/F(2.0-10.0~{\rm keV})$.\\
\end{flushleft} }
\label{spec_table}

\end{footnotesize}
\end{table*}

Insight-HXMT \citep{hxmt07, hxmt14} is the first Chinese X-ray space mission launched on 15$^{th}$ of June 2017.
It has three main detectors sensitive to different energy ranges: 
\textit{(i)} The low-energy detector (hereafter \textit{LE}) operates the Swept Charge Device 
sensitive to the energy range of $1.0-15.0$~keV whose effective area is 384~cm$^2$.
LE has two different field of view (FoV) options; the small FoV, 1.6$^\circ$~$\times$~6$^\circ$, and 
the big FoV, 6$^\circ$~$\times$~6$^\circ$.
There are also three full blocked detectors (also called blind detectors) on LE to estimate the internal gain and the background contribution.
\textit{(ii)} The medium-energy detector (hereafter \textit{ME}) consist of 1728 Si-PIN detectors
sensitive to $5.0-30.0$~keV energy band with the total effective area of 952~cm$^2$.
The FoV options of ME are the Small FoV of 1$^\circ$~$\times$~4$^\circ$, and 
the big FoV of 4$^\circ$~$\times$~4$^\circ$.
There are also three groups of blind detectors in which each group has 32 Si-PIN blind detectors to 
calculate the background components.
\textit{(iii)} The high-energy detector (hereafter \textit{HE}) made by 18 cylindrical NaI(Tl)/CsI(Na) PHOSWICH detectors
sensitive to $20.0-250.0$~keV with an additional blind detector. The FoV options for HE are 5.7$^\circ$~$
\times$~1.1$^\circ$ and 5.7$^\circ$~$\times$~5.7$^\circ$. The large total effective area of 5100~cm$^2$ 
in the hard X-ray band is the main advantage of Insight-HXMT. 

Aql X-1 is included in the observation list of Insight-HXMT as time of opportunity (ToO) source.
Unfortunately, when Aql X-1 underwent to the 2018 outburst, the source was unobservable by Insight-HXMT because of 
the criteria on the solar avoidance angle which must be greater than 70$^\circ$.
Aql X-1 has been observed by Insight-HXMT three times in March 2018
when the solar avoidance criteria has been passed and the source was still bright enough. Although the
observations do not cover whole outburst,  they are still valuable to study the spectral properties during the transitions to
quiescent state. the LE, the ME and the HE detectors were active simultaneously during observations and the exposure time was 10~ks for each pointing.
The total count rates before and after  the good time interval
corrections are listed in \autoref{spec_table} for each detector and each data set.
The data from the HE detector are not considered in the analysis because the source is background-dominated.

The data analysis has been done using the latest version of 
\ac{HXMTDAS} v2.0\footnote{\url{http://www.hxmt.org/index.php/dataan/}}.
We limited the pointing offset angle to 0.1$^\circ$ to avoid the slew data.
We constrained the maximum elevation angle of 12$^\circ$ for LE and 10$^\circ$ for ME.
Additionally for the LE detector, we bounded the elevation angle of bright earth as 40$^\circ$.
We set the maximum of geomagnetic cut-off rigidity to 8~GV for both LE and ME.
After we filtered the data using the criteria above, we obtained the spectra and the light curves by choosing the
small FoV for each detector (1.6$^\circ$~$\times$~6$^\circ$ for LE and 1$^\circ$~$\times$~4$^\circ$ for ME).
We also used the blind detectors to estimate the background spectra and light curves.
Using the blind detectors to estimate background level has been tested
with the blank sky observations by Insight-HXMT calibration team for the feasibility of this method. 
We created the response matrix files using the \textit{lerspgen} and \textit{merspgen} tasks in \ac{HXMTDAS} for transforming the channel number to energy. 

\begin{figure}[t]
\centering
  \includegraphics[angle=90, scale=0.35]{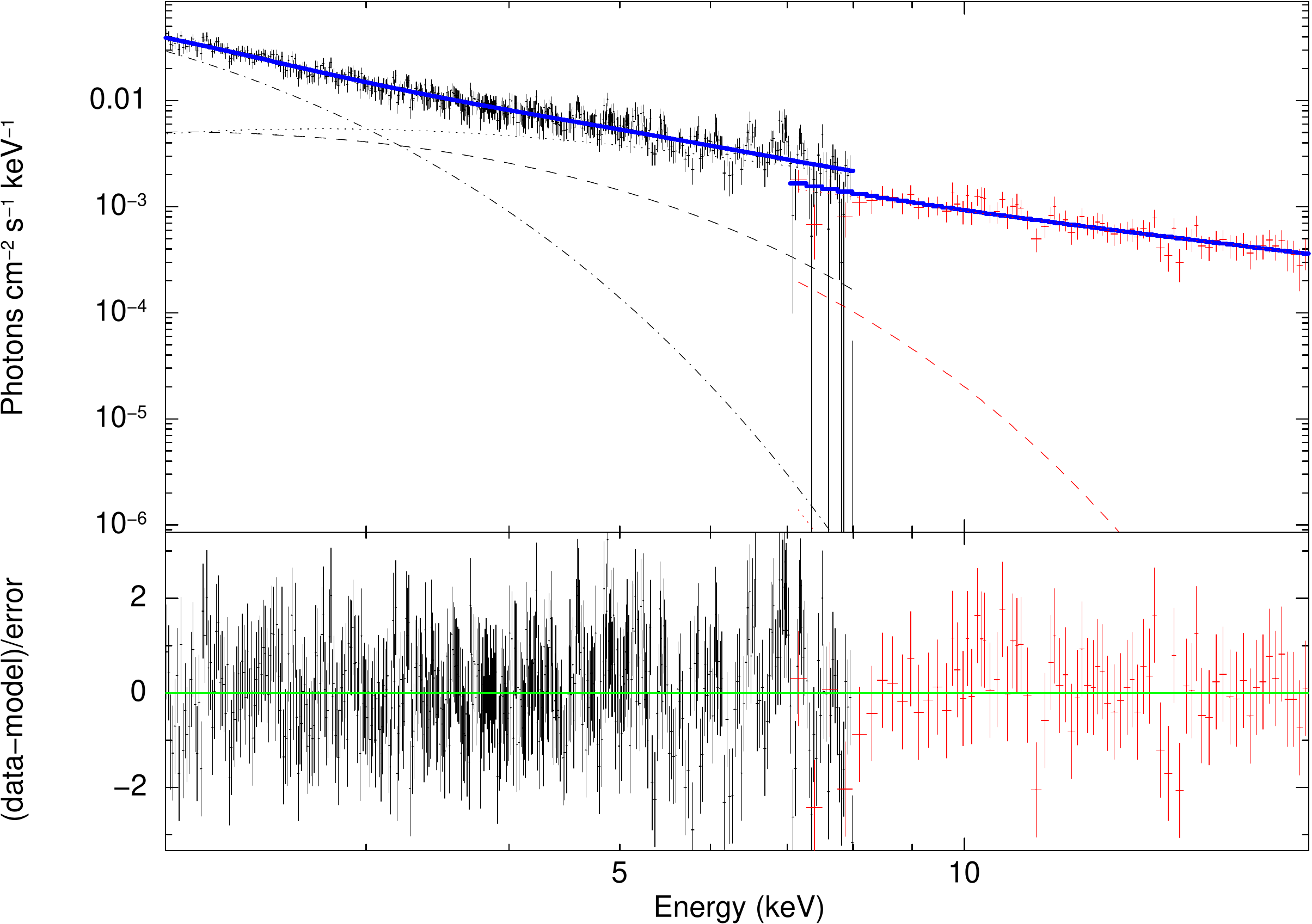}
     \caption{An example spectrum
     of Insight-HXMT observations of Aql X-1 in the energy range of $2.0-20.0$~keV , combination of LE (black) and 
     ME (red) detectors (ObsID O011466801001).
     The best fits of the \textit{bb+diskbb+compTT} model (\textit{V}) are shown with blue line. 
     Lower panel presents the residuals in terms of sigma.}
         \label{spectrum}
\end{figure}

We modelled the spectra by using the 12.10.0c version of XSPEC package\footnote{An X-ray spectral Fitting Package v12.10.0c, \url{https://heasarc.gsfc.nasa.gov/xanadu/xspec/}} \citep{xspec}.
We obtained spectra using each channel without grouping.
We used the energy range of $2.0-8.0$ keV and $7.0-20.0$~keV for the LE and the ME detectors, respectively.
We added 1.0\% systematic error to take into account
instrumental uncertainties \citep{chen2019}.
We applied \textit{phabs} as a photoelectric absorption model with the default cross-section in XSPEC version 12.0
\citep{verner96}.
The neutral hydrogen column density is fixed to $N_{\rm H}~=~3.4\times 10^{21}~{\rm cm}^{-2}$ \citep{mac03spec}.
After we loaded the LE data as group 1 and the ME data as group 2, we multiplied the whole model
with a constant value to calibrate the normalization levels of the detectors.
We fixed this constant to 1.0 for LE and the one for ME kept as the free parameter of the fit
while all other free model parameters are linked among LE and ME spectra.
We applied a set of models to represent the spectra; blackbody + power law (model I), 
blackbody + disk blackbody (model II), blackbody + comptonization (model III), disk blackbody + comptonization (model IV) 
and blackbody + disk blackbody + comptonization (model V).
We used \textit{compTT} in XSPEC as the comptonization model \citep{titar95}.
The plasma temperature of \textit{compTT} is fixed to $15.0$~keV \citep{gogus07} while the input soft photon (Wien) temperature is linked to the temperatures of the
blackbody components in model III and V assuming that the source of the up-scattered photons is the \ac{NS} poles,
and to the disk blackbody temperature in model IV.
In order to take into account the effectiveness of the iron line, we checked the quality of the models adding a Gaussian component
with fixed central line energy of $6.4$~keV and the line width of $0.6$~keV to each model.
Differently from RXTE data \citep{gungor+14, gungor+17b}, Gaussian component
was not statistically necessary.
We also obtained the hardness values for each data set using the ratio of the unabsorbed flux in the energy range of 
$2.0-10.0$~keV to the ones in the energy range of $10.0-20.0$~keV using the cflux task in XSPEC.
This energy ranges were chosen to be consistent with the hardness values obtained from \ac{MAXI} count rates.
We present an example spectrum of Insight-HXMT (ObsID: O011466801001) and the best fit of the model V in \autoref{spectrum}.
We list the output of our analysis in \autoref{spec_table}.
A detailed discussion about the physical interpretations of the adopted models is given in \autoref{spec-hxmt}.
Hereby, we must mention that the hard X-ray luminosity of Aql X-1 during Insight-HXMT
observation is too low and background dominates. So, we could not use the data from the HE detector.

\section{Discussion}
\label{discuss}

In order to better interpret the spectral output of Aql X-1 from Insght-HXMT during the decay stage,
we first studied the 2018 outburst according to the outburst morphology and the light curve properties.

\subsection{Classification of outbursts}

\ac{FRED} and \ac{LIS} type of outbursts  of Aql X-1 were first 
defined by \citet{mai08} based on the light curve morphology.
\ac{FRED} type outbursts have steep rising and exponential decay while
\ac{LIS} outbursts are the periods when the source is slightly more luminous than the quiescent state level ($\sim$~5$-$10 cnt/s in ASM; $\sim$~100 mCrab) with very high optical to soft X-ray flux ratio \citep{mai08}.
Aql X-1 showed a \ac{FRED} type outburst in February 2018, almost 190
days after the end of a \ac{LIS} type event started in  May 2017. This FRED type outburst reached its maximum 
on 26$^{th}$ of February 2018 in its X-ray light curve ($2.0-20.0$~keV).
\citet{spir18} reported the peak of Aql X-1 
in R band on 1$^{st}$ of March 2018 which is 3 days after its X-ray peak.

\citet{asa+12} reported different classification depending on the pattern of the relative intensity evolution
in the two energy bands below/above $15$ keV; slow-type (S-type)
and fast-type (F-type) outburst. The S-type outbursts have relatively longer ($\gtrsim$~9~days) initial hard-state,
while the  F-type outbursts have shorter one ($\lesssim$~5~days). The intensity in the energy range of $15.0-50.0$~keV
of the S-type outbursts reaches to its maximum
in the initial hard-state period and decreases dramatically at the hard-to-soft transition.
Differently, the intensity of the F-type of outbursts in the energy range of both $2.0-15.0$~keV and $15.0-50.0$~keV peaks after the transition.
Accordingly, the 2018 FRED type of outburst can be classified as S-type with its $\sim$10 days initial hard-state duration 
(see bottom panel of \autoref{fit-I}).

\begin{figure}[!t]
\centering
  \includegraphics[angle=0,width=0.47\textwidth]{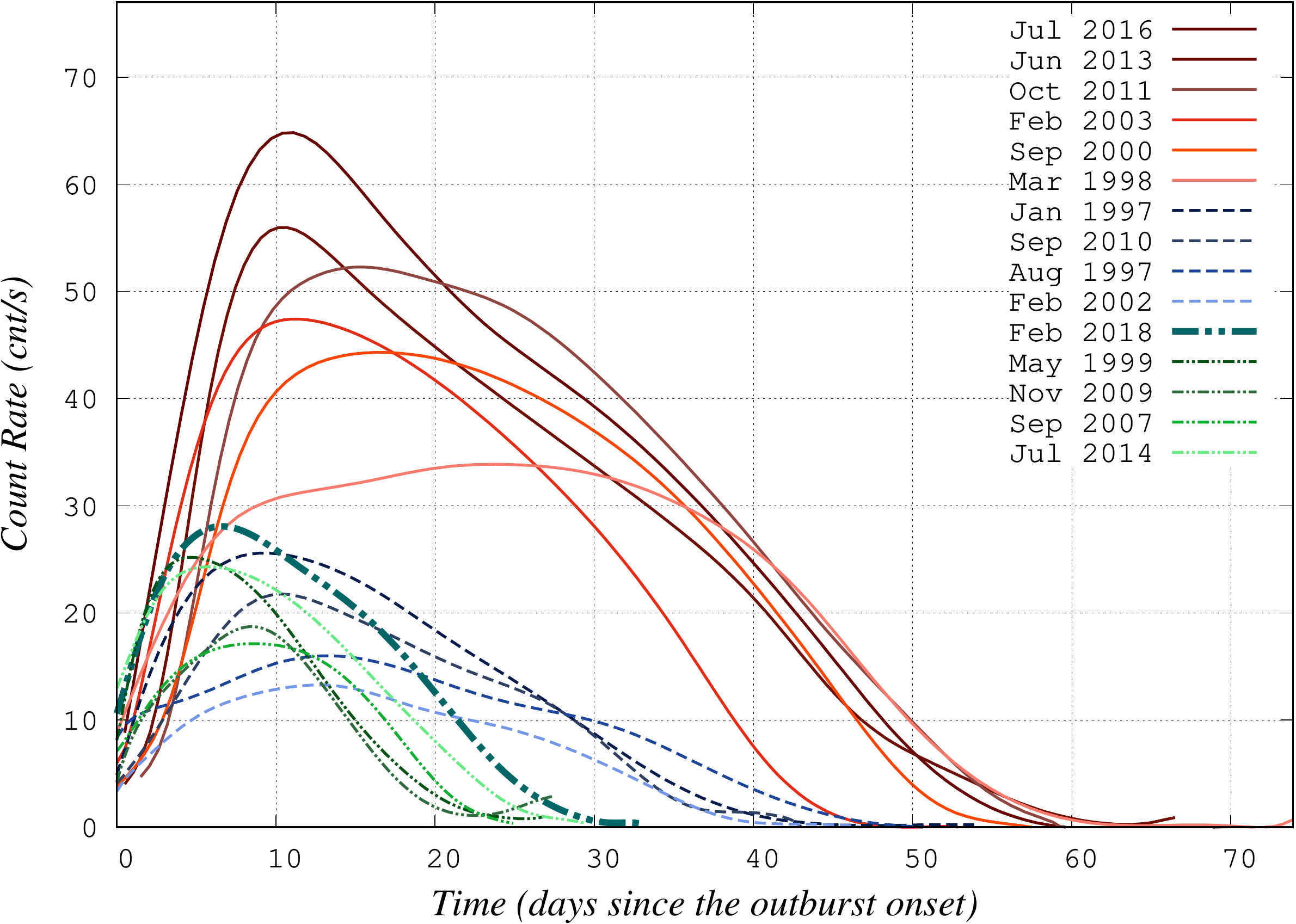}
     \caption{The light curves of the \ac{FRED} type outbursts of Aql X-1 calibrated to the beginnings.
     The light curves are smoothed for easy visualisation.
     This graph is the revised version of the 
     one presented by \citet{gungor+17a} by adding the 2018 outburst.
     The labels are sorted descendingly according to the peak count rate of each outburst. The long-high, the medium-low
     and the short-low outbursts are shown with different tones of red, blue and green, respectively.  
     }
         \label{light}
\end{figure}

Another type of classification was presented by \citet{gungor+14}, 
based on the peak count rate and the duration of the \ac{FRED} type outbursts. 
This work excluded the \ac{LIS} type outbursts since they do not have a systematic outburst pattern.
In this study, three outburst types --long-high, medium-low and short-low-- were defined 
and the main physical mechanism to discriminate
outburst types is mentioned as the irradiation.
Following their methodology, we first smoothed the $2.0-20.0$~keV \ac{MAXI} light curve of 
the 2018 outburst with B\'ezier spline method and we re-scaled the times to the outburst onsets.
We, then, compared it with all of the other outbursts that Aql X-1 showed (\autoref{light}).
Hereunder, the 2018 outburst is classified as short-low type with the peak count rate of $\sim$30~cnts$^{-1}$
and the outburst duration of $\sim$30~days, 
although its rising stage is slightly milder than the other short-low type outbursts.

\subsection{Partial accretion in the propeller stage}
\label{prop}

The decay stages of the light curves of the \ac{FRED} type outbursts 
in \ac{LMXB} systems generally have two different decay trend --the slow and the fast decay stages--.
The critical \textit{cut-off} point, also called \textit{knee}, between these two stages is mentioned
as the possible transition from the accretion stage 
to the weak propeller stage \citep{zha+98pro, gil+98, cam+98, ibr09, asai+13}.
Herein, we have to mention that these \textit{cut-off} behaviours have been observed from the light curves of 
both NS-LMXBs and BH-LMXBs in which the propeller effect is not expected from the systems 
with BHs since they do not have magnetic fields.
This cut-off in \acp{LMXB} can also be explained via the thermal disk instability model \citep{las01}
which is the only scenario for BH-LMXBs while for NS-LMXBs the transition from accretion to propeller stage is an alternative.
If the reason of the existence of the knee at the decay stage of Aql X-1 is the same as in BH systems,
the technique of calculating the mass fraction rate in the weak propeller stage using the knee would be invalid.

\begin{figure}[!t]
\centering
  \includegraphics[angle=0,width=0.47\textwidth]{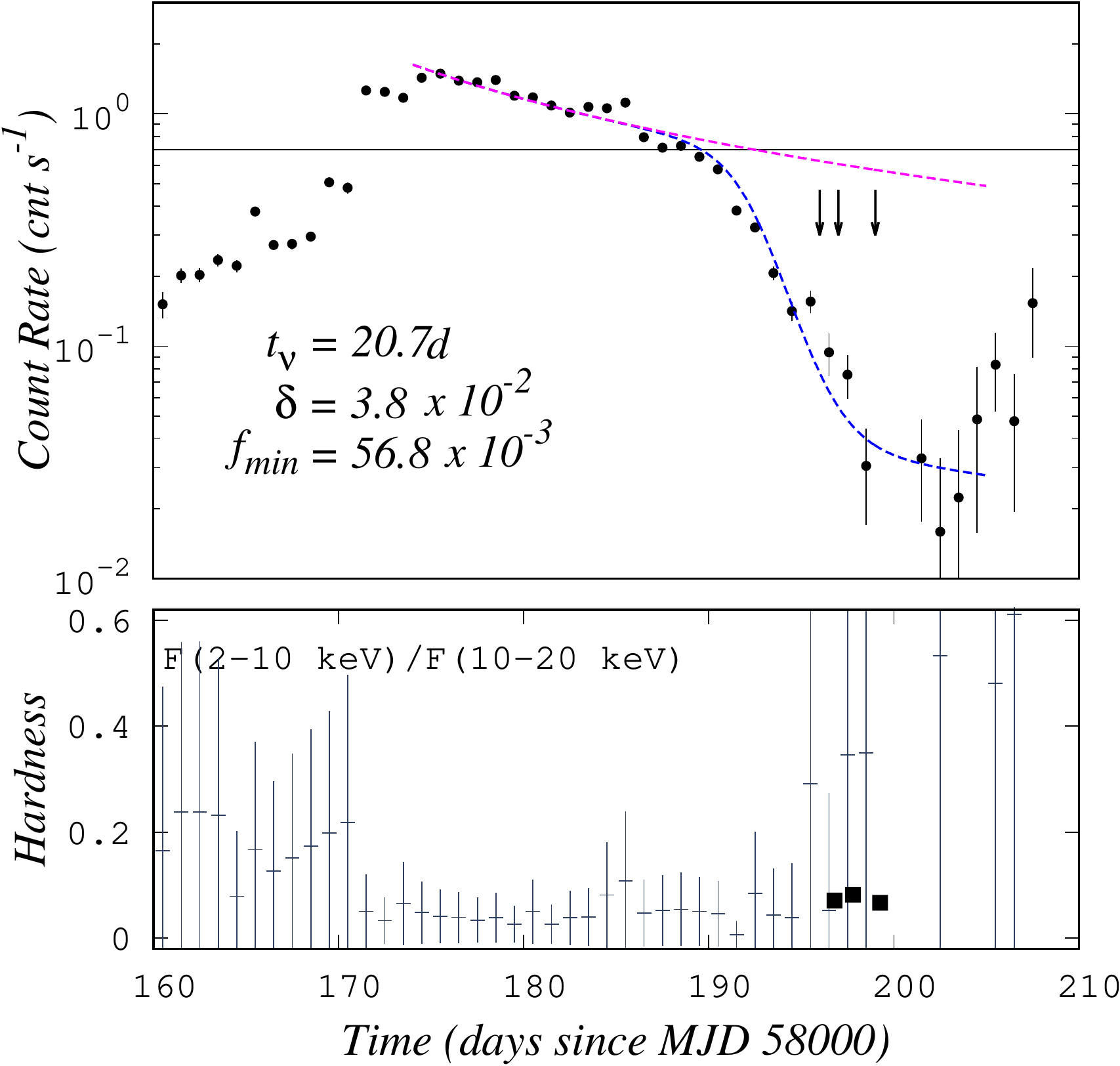}
     \caption{\textit{Upper panel:} The \ac{MAXI} light curve of the 2018 outburst of Aql X-1 with the best fit to the model.
     The pink and the blue curves represent the best fit of \autoref{luminosity_acc}
     to the slow decay stage and $L(t)$ to the whole of the light curve, respectively.
     The vertical dashed line shows the critical count rate level of the source while propeller transition.
     The upside-down arrows show the observation times of Insight-HXMT.
     \textit{Bottom panel}: The hardness evolution along the 2018 outburst obtained from the ratio of
     the \ac{MAXI} count rates in the energy ranges of $2.0-10.0$~keV and $10.0-20.0$~keV.
     The hardness values obtained from Insight-HXMT data are shown with black filled squares.
     }
         \label{fit-I}
\end{figure}

A unique technique, using the X-ray light curve, has been proposed by \citet{gungor+17b}
to calculate the mass accretion rate in the propeller stage as a function of the fastness parameter.
The origin of the X-ray luminosity, is the gravitational potential energy of the transferring 
material,
\begin{equation}
\label{lumin_x}
L_{\mathrm X} = \frac {G M_{\ast} \dot{M}_{\ast}}{R_{\ast}}, 
\end{equation}
where $G$ is the gravitational constant, $M_{\ast}$ and $R_{\ast}$ are the mass and the radius of the \ac{NS}, and 
$\dot{M}_{\ast}$ is the mass accretion rate falling onto the poles of \ac{NS}.
Accordingly, the X-ray luminosity is directly proportional to the mass accretion rate ($ L_{X} \propto \dot{M}_{\ast}$).
Assuming that the slow decay corresponds to the accretion stage,
all of the material reaching the inner layers  of the disk ($\dot{M}$) is accreted, therefore,
the exponential trend of the slow decay stage gives the time variation of $\dot{M}$. The slow decay can be represented 
by using the function \citep{gungor+17b};
\begin{equation}
\label{luminosity_acc}
{L}(t)= {L}_{0} \left(1 + \frac{t-t_0}{t_{\nu}} \right)^{-\alpha},
\end{equation}
where t$_{0}$ is the time of the peak,
L$_{0}$ is the luminosity at the time of t$_{0}$, and t$_{\nu}$ is the free fit parameter proportional to the timescale of the outburst. 
$\alpha$ is the power$-$low index related to the pressure and the opacity in the disk. 
We used $\alpha$ of 1.25 for the fit in which this value is suitable for a gas pressure dominated disk with bound free opacity \citep{can+90, eks11}.
This formula also gives $\dot{M}(t)$ from \autoref{lumin_x}.
It is assumed that the time variation of $\dot{M}$ follows the same 
trend in the weak propeller stage in a condition of that time variation of the mass transfer rate throughout the disk continues with the same trend.
Thus, the mass transfer ratio of the falling material onto the \ac{NS} and the material reaching to the inner layers, $f \equiv \dot{M}_{\ast}/\dot{M}$, can be obtained via the 
ratio of the observed luminosity in the fast decay stage and the calculated luminosity from the \autoref{luminosity_acc} for the 
corresponding time. 
$f$ function is expected to be a step function in the simplest picture of an ideal propeller
surrounded by an infinitely thin disk. If the disk has a scale
height, the accretion can proceed from higher latitudes of the disk \citep{rom04, eks11}, then, one can expect $f$ as a 
smoothed step function;
\begin{equation}
f = \frac12 \left[ 1 + f_{\min} + (1-f_{\min}) \tanh \left(\frac{\omega_c - \omega_\ast}{\delta}\right) \right],
\label{step}
\end{equation}
where $f_{min}$ is the bottom level of the step function, $\omega_\ast$ and $\omega_c$ are the fastness parameter and its critical value for the propeller transition,
and $\delta$ is the value of the smoothness. So, the $\delta$ parameter is related to the thickness of the inner layer of the accretion disk
while $f_{min}$ is linked with the extra luminosity sources other than accretion such as NS cooling in the quiescent level.
Since \autoref{luminosity_acc} represents the slow decay and \autoref{step} is a smoothed step function, multiplication of these two equations can represent the whole decay.
Before the knee, step function equals to $1$ and only \autoref{luminosity_acc} is valid while after the knee $L_X$ goes to quiescent level smoothly.
In this study, we applied the technique to the \ac{MAXI} light curve of the 2018 outburst of Aql X-1 and obtained the free parameters in the model (\autoref{fit-I}).
We estimated $t_\nu$ of 20.7{$\pm$}1.3~d, $\delta$ of (3.8{$\pm$}1.1)$\times$10$^{-2}$ and $f_{min}$ of (56.6{$\pm$}13.0)$\times$10$^{-3}$. The estimated 
value of $\delta$ is
consistent with the results of previous outbursts presented by \citet{gungor+17b}.
The method allows us to estimate the fastness parameter for a given time using the relation below;
\begin{equation}
\omega_{\ast}(t) = \left[ \frac {L(t)}{L_c}\right]^{-3/7}
\label{fast2}
\end{equation}
where L$_c$ is the critical luminosity at the time of the transition from the accretion stage to the propeller stage (see \citealp{gungor+17b} for derivation).
The calculated fastness parameters for the times of Insight-HXMT observations are given in \autoref{spec_table}.
The given fastness parameters show that the angular velocity of the star is larger than the Keplerian angular velocity in the inner radius of the disk and the material repelled by the centrifugal barrier from the inner layers of the disk to larger radii.

\subsection{Spectral output of Insight-HXMT}
\label{spec-hxmt}

The first model (model I) is the combination of a blackbody and a power law.
We assume that the blackbody represents the thermal radiation from the source and the power law component represents the non-thermal radiation (if any).
While, this model gives relatively good reduced chi-square ($\chi^2/d.o.f.$) values ($\sim$~0.87),
the blackbody temperatures are very high and evolve to higher values for lower flux levels
while the photon indexes are around $\sim$~2.5 which match with
previous results of \acp{LMXB} \citep[e.g.][]{rem+06, seifina+15}
This situation shows that the model must be modified since the blackbody temperatures can not be explained via physical processes even if the data is quite good modelled.

In model II, we implemented a combination of models of blackbody, for the radiation from \ac{NS} itself 
and disk blackbody for the contribution of the inner layers of the accretion disk. The temperatures of the blackbody components 
are also quite high
as seen in model I, although the model represents the data well with the $\chi^2/d.o.f.$ of $\sim$~0.90. 
The normalization of the disk blackbody model is defined as $(R_{in}/D_{10})^2\cos\theta$, where R$_{in}$ is 
the inner disk radius in km,
D$_{10}$ is the distance in the unit of 10~kpc and $\theta$ is the angle of the disk.
The normalization value for the  observation of ObsID P011466801001 is 40 which gives R$_{in} = 2.81$~km/$\sqrt{\cos\theta}$ by taking the source distance as 4.5~kpc \citep{gal+08}.
This value can only give a reasonable value for angles really close to 90$^\circ$ (edge on view).
Given the fact that this model and the previous one give an unsatisfactory description, we need to add an extra model to take into account comptonization which is expected for \acp{LMXB} in low accretion regime.

We checked the effect of the up-scattered photons from the Compton cloud by modelling the spectra using the combination of blackbody and comptonization (model III).
This model can also represent the spectra statistically good with $\chi^2/d.o.f.$ of $\sim$~0.90.
The blackbody temperature throughout this model evolves from $0.37$~keV to $0.69$~keV. This temperatures are lower than the ones obtained from model I and II.

We, then, added the comptonization model to diskbb component (model IV) in which the seed photon temperature of the comptonization model is now linked
to the temperature of the inner radius of the disk in diskbb component. The output temperatures of model IV increase in time along three data sets.
This is not an expected phenomena since the inner radius of the disk shrinks back along time at the decay stage of an outburst.
The normalization of the diskbb component of the first observation set is $\sim$~500 which results
the inner radius of the disk is R$_{in} = 10.06$~km for face-on viewing angle ($\theta=0$) for the source distance of 4.5~kpc.

We, lastly, implemented a model which is a combination of a blackbody, a disk blackbody and a comptonization (model V) 
with the seed photon temperature linked to the blackbody temperature assuming that the up-scattered photons are coming from the \ac{NS} poles.
Similar to the other models the spectra are well fitted with an average $\chi^2/d.o.f.$ of $\sim$~0.88. On the other hand, we do not see the high
blackbody temperatures seen in model I and the normalization of the disk blackbody component is similar to the one obtained from model V which gives a reasonable inner radius of the disk.
The F statistic value and its probability of adding the comptonization component are $50.33$ and $1.79\times10^{-21}$, respectively. Low probability and large F$-$test value shows that adding the third component provides a significant improvement.
Ultimately, the most reliable model overall these 5 models is the model V with both its physical interpretation and 
the acceptable model parameters.

In order to compare spectral output of our analysis, we first found the data with similar fastness parameters in the literature.
The fastness parameters of \ac{RXTE} data with the ObsId of 50049-03-04-00, 96440-01-09-05 and 96440-01-09-12
are 1.06, 1.06 and 1.07, respectively \citep{gungor+17b}.
These three observations were acquired
while the source was transiting from the soft-high state to 
the hard-low state,
while the first belongs to the 2000 outburst and the rest belongs to the 2011 outburst.
\citet{gungor+17b} mentioned that the temperature values without comptonization model are too high to be explained via physical processes
even if the model may fit the spectrum mathematically well with reasonable $\chi^2/d.o.f.$ values.
We see the same situation in our analysis. This strengthens the method created by \citet{gungor+17b}.
in which the method can be also used to compare the data in different luminosity levels and the data obtained by different space missions since it is based mainly on using unitless parameters ($\omega_{\ast}$ and $f \equiv \dot{M}_{\ast}/\dot{M}$). Parameters listed in \autoref{spec_table} show that adding comptonization corrects the high temperatures in model I and model II. 
This implies that the electron cloud is effective in low accretion rates.
 
\section{Conclusions}
\label{conc}

We present the output of the spectral analysis of three Aql X-1 observations obtained by Insight-HXMT.
We model the spectra in the energy range of $2.0-20.0$~keV combining the data from the LE and the ME detectors
with a set of models.
We compared our results to the ones obtained by \ac{RXTE} data with similar fastness parameters.
Differently from \ac{RXTE}, Insight-HXMT is able to cover
a broader energy range combining LE and ME detectors including energies below $3.0$~keV and with its
better CCD type energy resolution.
We show that the temperatures of the blackbody components are very high to be resulted from physical processes 
for model I and model II, which demonstrates that these models do not well work in low luminosity/accretion regimes.
A comptonization component to take into
account the inverse Compton process of up-scattered photons, takes the high blackbody temperatures to reasonable values.
This indicates the existence of electron cloud between the inner disk and \ac{NS} in the low accretion regime.

We study the 2018 \ac{FRED} type outburst. We show that the 2018 outburst is a member of 
S-type according to the classification presented by \citet{asa+12} and belongs to the short-low class
according to the classification of \citet{gungor+14}.
Applying the technique presented by \citet{gungor+17b},
we show that the Insight-HXMT observations are just after the 
transition from the accretion stage to the propeller stage.
The hardness parameters obtained from MAXI data is unstable at the time of Insight-HXMT observations. But on the other hand, the ones throughout our
Insight-HXMT analysis are consistent with the previous trend of MAXI, 
indicating that the system is still in the high-soft regime or in the transition 
from high-soft state to low-hard state.

\section*{Acknowledgements}
We thank anonymous referee for detailed review and constructive comments.  CG acknowledges support from Chinese Academy of Sciences President's 
International Fellowship Initiative (CAS/PIFI).
This study has made use of the data from the Insight-HXMT mission, a project funded by China National
Space Administration (CNSA) and the Chinese Academy of Sciences (CAS).
This research has made use of the MAXI data provided by RIKEN, JAXA and the MAXI team and the results 
provided by the ASM/RXTE teams at MIT and at the RXTE SOF and GOF at NASA’s GSFC.
This work is supported as well by the National Key R\&D Program
of China (2016YFA0400800) and the National Natural Science
Foundation of China under grants 11473027, 11733009,
U1838201, U1838202 and U1838104.

\end{document}